# Outer-Valence Intermolecular Coulombic Decay in Hydrogen-Bonded Complexes Induced by Resonant Two-Photon Excitation by 4.4 eV Photons

Namitha Brijit Bejoy,[1] Nitin. K Singh,[2] Balanarayan Pananghat[2] and G. Naresh Patwari[1]*

[1] Department of Chemistry, Indian Institute of Technology Bombay, Powai Mumbai 400076 India. E-mail: naresh@chem.iitb.ac.in

[2] Indian Institute of Science Education and Research (IISER) Mohali, S. A. S Nagar, Mohali 140306 Punjab, India.

**Abstract**

Photoexcitation of van der Waals complexes can lead to several decay pathways depending on the nature of the potential energy surfaces. Upon excitation of a chromophore in a complex, ionization of its weakly bound neighbour via energy transfer happens via a unique relaxation process known as intermolecular Coulombic decay (ICD); a phenomenon of renewed focus owing to its relevance in biological systems. Herein, we report an experimental evidence of outer-valence ICD induced by multiphoton excitation by near UV radiation of 4.4 eV photons, hitherto unknown in molecular systems. In the binary complexes of 2,6-difluorophenylacetylene with aliphatic amines such as dimethylamine and trimethylamine a resonant two-photon excitation is localized on the 2,6-difluorophenylacetylene chromophore. The absorbed energy is then transferred to its hydrogen-bonded partner amine and resulting in the formation of an amine cation. The kinetic energy distribution of the amine cations is invariant with respect to the initial geometry of the binary complex. The present experimental results combined with electronic structure calculations provide valuable insights into the nature of ICD in van der Waals complexes and more importantly, the role of ICD as a fast, efficient, and prominent decay channel following excitation at modest (4.4 eV) photon energies.

**Introduction**

The ability of a chromophore to absorb energy and to ionize atoms/molecules embedded within its van der Waals sphere of interaction, via energy transfer is referred to as Intermolecular Coulombic Decay (ICD).[1] ICD is an ultrafast and efficient bound-to-continuum energy transfer process, resulting in the ejection of low kinetic energy electrons.[2] The initial theoretical and experimental investigations on the ICD phenomenon were carried out primarily on $(HF)_n$ clusters,[1] rare gas dimers,[3–5] water clusters,[6,7] wherein the ICD process leads to the formation of two charged fragments resulting in a Coulomb explosion. Over the years, ICD has been observed in a variety of systems,[1] viz., ammonia clusters,[8] benzene clusters,[9] mixed alkali dimers,[10] and more recently in endohedral fullerenes.[11] A key experimental signature of the ICD phenomenon is the appearance of low energy secondary electrons,[12,13] which can initiate subsequent structural modifications in biomolecular systems.[14] The ICD as a decay channel upon photoexcitation has been conceived to be a prominent source of such secondary electrons leading to the repair of DNA lesions.[15] The ICD process following electron impact ionization of the inner valence electron of water molecule in a hydrogen-bonded, biologically-relevant, water-tetrahydrofuran (THF) complex, results in a Coulomb explosion due to ICD mediated generation of water and THF cations.[16] However, in the case of protic systems such as water/ammonia clusters the proton transfer process competes with the ICD.[17,18]

The process of ICD originating out of excitation of outer-valence electrons is unknown in molecular systems, even though it has been recognized that it is a very general phenomenon that occurs independently of the exact excitation route.[1] The photoexcitation of outer-valence electrons in atomic and molecular clusters, affected by near UV radiation, can also initiate ICD through a resonant multi-photon absorption process. For instance, photoexcitation of a heterodimer complex [A···B] with near UV radiation resonant with the electronic states of chromophore A, can result in the ionization of B via energy transfer. In this work, we demonstrate the ICD process ensuing from an outer-valence excitation in binary complexes of 2,6-difluorophenylacetylene (DFPHA) with dimethylamine (DMA) and trimethylamine (TMA), as illustrated by the mechanism depicted in Figure 1,[15] and by measuring the kinetic energy of the amine fragments using a velocity map imaging technique

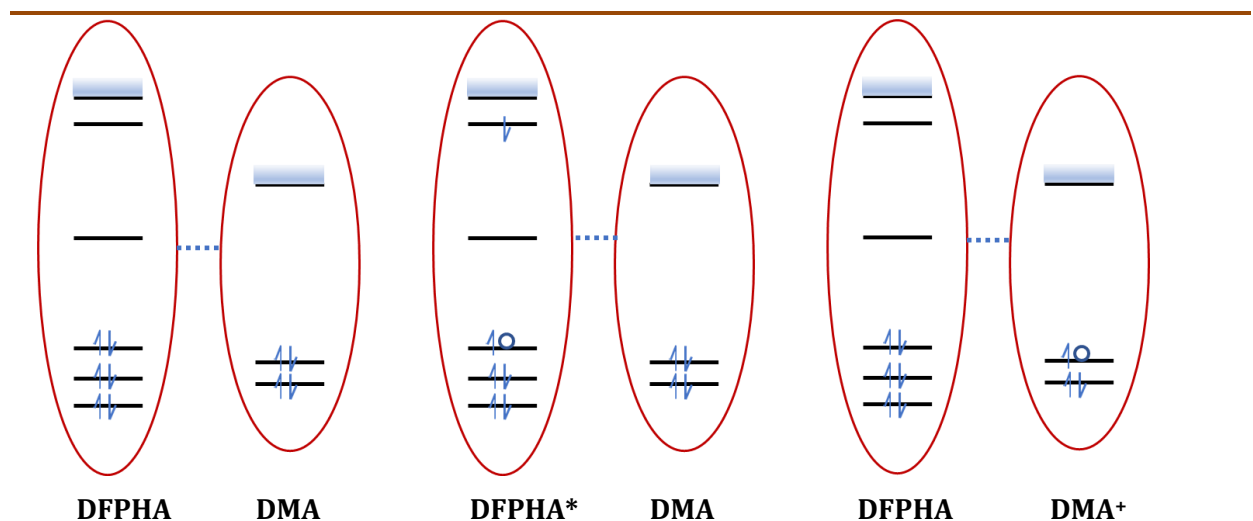

**Figure 1.** A cartoon representation of the mechanism of ICD process starting from the DFPHA···DMA complex in the neutral ground state (left panel), an initial resonant two-photon absorption leads to an excitation localized on the DFPHA moiety well above the ionization threshold of DMA (centre panel), the ICD process ionizes DMA leading to the formation of DFPHA···[DMA]$^+$ (right panel), which results in a detectable DMA$^+$ due to repulsive nature of the potential energy surface.

following a resonant two-photon excitation.

**RESULTS and DISCUSSION**

The electronic spectrum of the DFPHA chromophore, in the near UV region (4.4 eV), is weakly perturbed by its van der Waals association with the amines as shown in Figure 2. The interaction of DFPHA chromophore with the two amines DMA and TMA leads to a marginal perturbation of its electronic excitation by about tens of milli-eV. Presented in Figure 2 is the electronic spectrum of the DFPHA monomer recorded using fluorescence detection technique, due to its inability to undergo resonant one-color two-photon ionization since its adiabatic ionization potential (9.11 eV) is higher than the available energy (8.86 eV) in the experiment.[19] The electronic spectra of the binary complexes of DFPHA with amines (DFPHA-DMA and DFPHA-TMA) were recorded as action spectra by monitoring the amine cation (DMA$^+$/TMA$^+$), arising from the excessive fragmentation following a resonant two-photon ionization process. The electronic spectra of the complexes show marginal shifts (less than 30 meV) in the bands which indicates that the electronic excitation is localized on the DFPHA chromophore. Moreover, the electronic spectra of DFPHA complexes with DMA and TMA recorded using the laser induced fluorescence method yield identical spectra, as shown in Figure S1 (see the Supplementary Information).[20] The fluorescence spectra also negate the possibility of excited state electron transfer process, which would lead to fluorescence quenching. The nature of the interaction between the chromophore and the two amines was probed using the IR-UV double resonance spectroscopy method and the resulting IR spectra are shown in Figure S2 (see the Supplementary Information). The results were interpreted using *ab-initio* calculations, leading to the structures with $C_{AC}$–H···N hydrogen bonding ($C_{AC}$ refers to the terminal carbon atom of the acetylenic group) for the C1 and C2 complexes and incorporating the lone-pair (Lp)···π interaction for the C3 complex as shown in Figure 2.[19] Observation of amine cations ($DMA^+$/$TMA^+$) following resonant two-photon excitation of the DFPHA chromophore is a consequence of the ICD process. The one-dimensional potential energy curves (PECs) for the $C_{AC}$–H···N hydrogen-bonded C1 and C2 complexes along the $C_{AC}$–H···N distance, shown in Figure 3, clearly indicate the repulsive nature of the ground state for DFPHA···DMA and DFPHA···TMA complex cations. The repulsive nature of these PECs is reconciled by the fact that the entire positive charge is carried by the amine fragment in the complex cation and converges to the asymptote consisting of neutral DFPHA and the DMA cation, which is in accord with the observed fragmentation pattern. The observation of the amine cations at the asymptote is due to the lower ionization potential of the amines (DMA 8.32 eV and TMA 7.82 eV) in comparison to DFPHA (9.11 eV).[19] The kinetic energy of the cationic amine fragments following the ICD process were measured using velocity map imaging (VMI) technique,[21] which resulted in isotropic images, shown in Figure 3,[10] and are attributed to the longer dissociation time relative to the rotational reorientation.[10,22] The total kinetic energy release (TKER) in the DMA complex has a broader distribution (maximum around 0.03 eV) than the corresponding TMA complex (maximum around 0.02 eV) as evident from the ion images. The TKER for the two different structural isomers C2 and C3 of TMA are almost identical, which can be attributed to the same dissociation asymptote, even though the initial geometry is different.

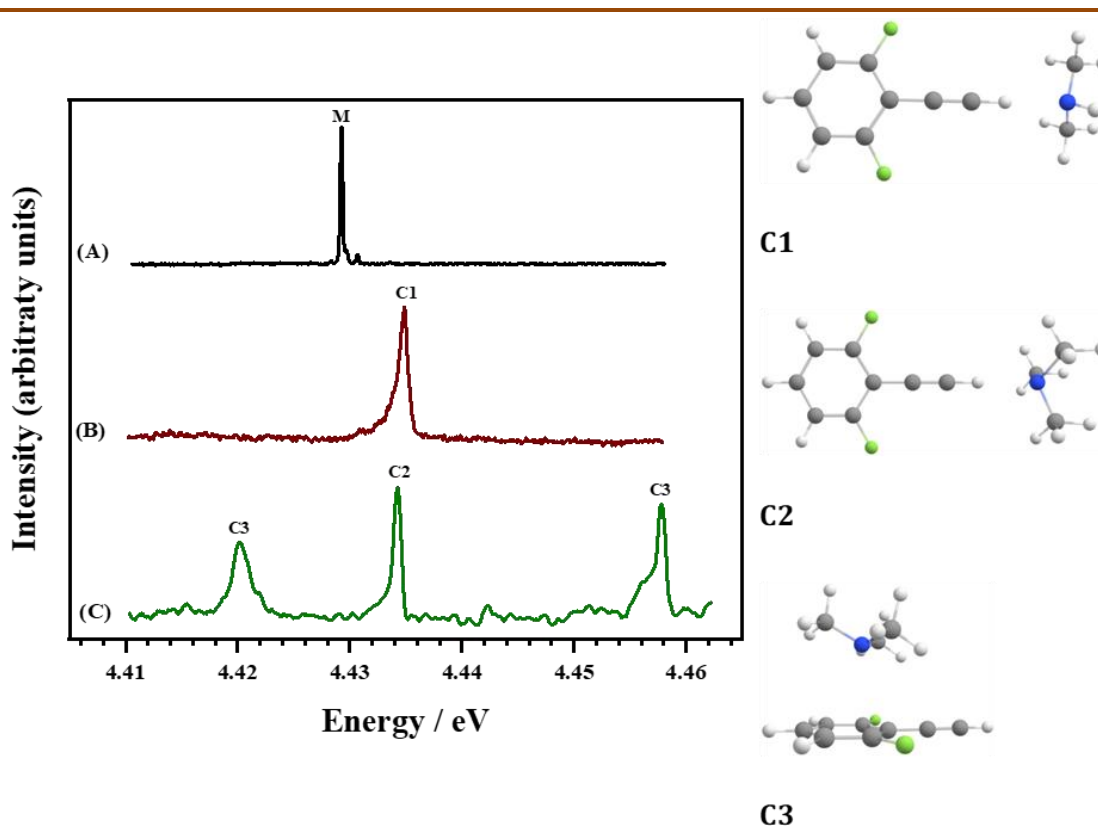

**Figure 2.** The electronic spectra of the isolated DFPHA (trace-A) and its complexes with DMA (trace-B) and TMA (trace-C). The electronic spectrum of the DFPHA monomer was recorded using laser induced fluorescence method owing to its inability to undergo resonant two-photon ionization as its adiabatic ionization potential is higher than 8.86 eV. On the other hand, resonant two-photon ionization of the amine complexes yields amine cations, which were monitored to record the electronic spectra as action spectra. The pointer 'M' indicates monomer, while 'C1', 'C2', and 'C3' indicate the various binary complexes with amines. The structures of the binary complexes, shown in the right panel, were determined via IR-UV double resonance spectroscopic technique in combination with *ab-initio* calculations.

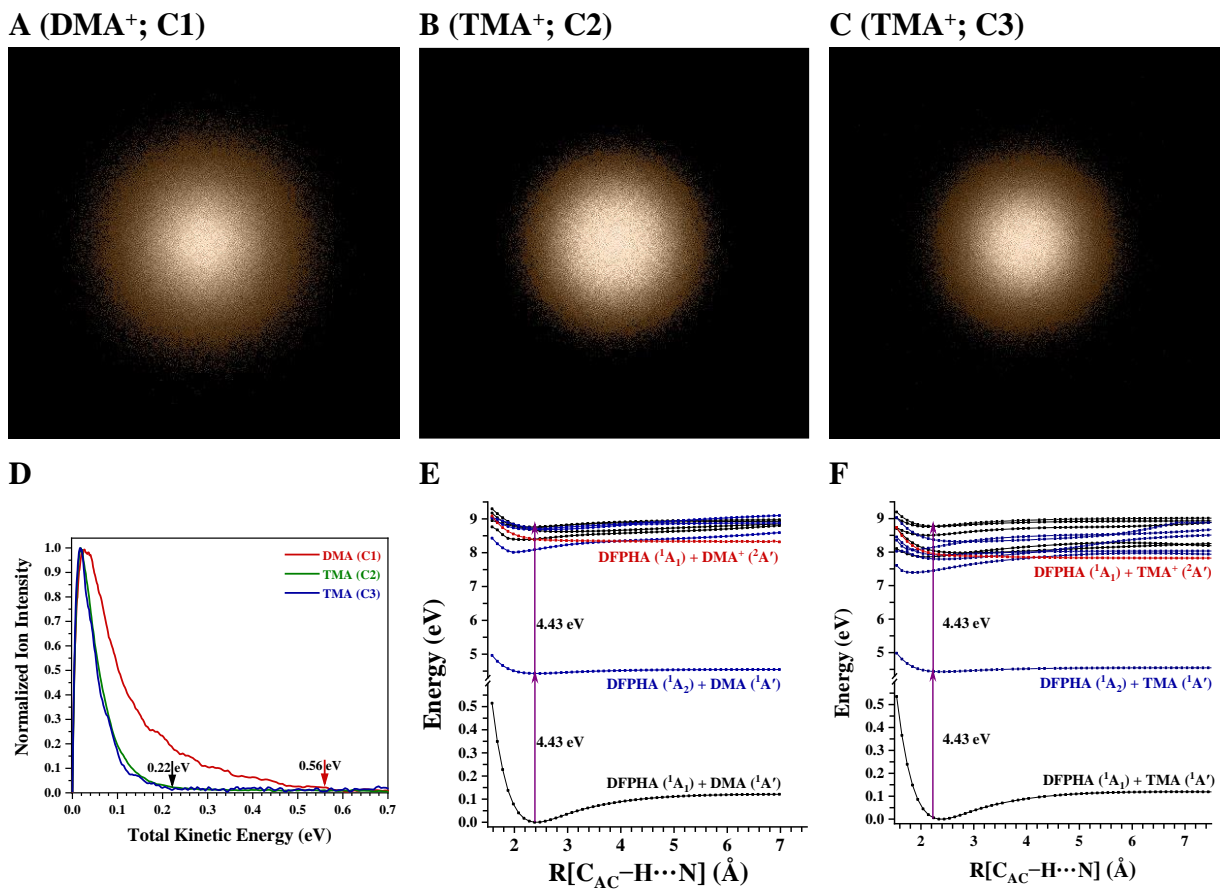

**Figure 3.** Velocity map raw images of the amine cations (A) DMA$^+$; C1, (B) TMA$^+$; C2, and (C) TMA$^+$; C3 originating out of the binary complexes with DFPHA. (D) The plots of total kinetic energy release for the three amine cations were extracted following Abel-transformation using BASEX method. Note that the kinetic energy profile for the DMA$^+$ is broader and distinctly different from that of the TMA$^+$, while the kinetic energy profiles for TMA$^+$ originating from two structurally distinct initial geometries are almost identical. The maximum values for the total kinetic energy are about 0.56 and 0.22 eV for the DMA$^+$ and TMA$^+$, respectively, indicated by arrows. (E and F) Relaxed one-dimensional potential energy curves (PECs) for the ground and excited states of binary complexes of DFPHA with DMA and TMA, respectively, along the C$_{AC}$–H···N hydrogen-bonded intermolecular distance calculated at the MCSCF/aug-cc-pVDZ level. The A' and A'' states are shown in back and blue, respectively, represent the potentials of the ground and the excited states of the neutral complex, whilst the red curve depicts the potential energy curve for the ground state of the complex cation. The exit channels for the ground, the first excited state of the neutral complex, and the complex cation ground state are shown. The PEC for the complex cation in its ground state, (red curve) is repulsive and produces a neutral ground state of DFPHA ($^1A_1$ state) and the ground state of DMA$^+$ / TMA$^+$ ($^2A'$ state).

The electronic excitation of the DFPHA chromophore to high energy excited states, albeit below its adiabatic ionization potential, initiates the relaxation process in the presence of a weakly interacting DMA, resulting in the transfer of the excess energy to its neighbour, ejecting an electron in this process results in [DFPHA-DMA]⁺ complex cation. This mechanism for the production of the [DFPHA-DMA]⁺ complex cation is ICD. ICD has been known to generally occur after the ionization of an inner-valence-shell electron of an atom/molecule enclosed in an environment.[5] From experimental observations, in the present scenario, the ICD occurs due to excitations from outer-valence electrons of DFPHA. Even though the ICD process occurring following excitation of inner-valence electrons has been well established in serval cases, it has been rarely observed following outer-valence electronic excitation. The theoretical explanation for the experimental observation of the DMA cation (DMA$^+$) is explained by electronic structure calculations with the aid of relaxed PECs of the DFPHA⋯DMA and DFPHA⋯TMA heterodimers as a function of the variation in the hydrogen bond length, R[$C_{AC}$–H⋯N], depicted in Figure 3, which gives the information regarding the exit channels for some of the states.[22] A resonant two- photon excitation of around 8.86 eV, owing to an intermediate first excited state at around 4.43 eV, results in a Frank-Condon transition to a plethora of states which are marginally above the single ionization threshold (red curve), but localized on the DFPHA moiety. Thus, the initial excitation to electronic states around 8.86 eV relaxes via the ICD process knocking off an electron from DMA/TMA, resulting in the formation of DMA$^+$/TMA$^+$ as illustrated in Figure 1. The PECs for the ground state of [DFPHA⋯DMA]$^+$ and [DFPHA⋯TMA]$^+$ complex cations (red curves in Figure 3E and 3F) are repulsive with the exit channel leading to the neutral ground state of DFPHA ($^1A_1$ state) and ground state of DMA$^+$ /TMA$^+$ ($^2A'$ state) at the asymptote. Further, the PECs of the ground state of the [DFPHA⋯DMA]$^+$ and [DFPHA⋯TMA]$^+$ complex cations suggest a total kinetic energy release of 0.54 and 1.04 eV in the centre-of-mass frame for the dissociation of the corresponding complex cation starting from the neutral ground state equilibrium geometry. The total kinetic energy release traces (Figure 3D) indicate the maximum values for the appearance of DMA$^+$ and TMA$^+$ are about 0.56 and 0.22 eV, respectively. In the case of DMA$^+$ the experimentally observed maximum kinetic energy value (0.56 eV) is in excellent agreement with the asymptotic value of 0.54 eV (within the experimental error), which suggest that the initial excited two-photon state directly couples with the ground state of the [DFPHA⋯DMA]$^+$ complex cation. On the other hand, the maximum kinetic energy (0.22 eV) observed for the TMA$^+$ is much less than the asymptotic value of 1.04 eV indicating the involvement of other excited states in the de-excitation processes.

Summarizing, we have demonstrated the ICD process following resonant two-photon excitation of 2,6-difluorophenylacetylene in its weak hydrogen-bonded complexes with alkyl amines at modest photon energies (around 4.4 eV). For structural isomers, the kinetic energy release was found to be independent of the initial geometry as both isomers converge to the same asymptote. The present set of experiments demonstrates that the presence of chromophores that absorb in the near UV region can under appropriate conditions result in the ICD process, and are likely to be relevant in the biological milieu in addition to the production of secondary electrons. Based on the present set of results, ICD for the chromophores that absorb in the near UV region is possibly more prevalent than what is currently realized.

**Acknowledgments**

This material is based upon work supported in part by Science and Engineering Research Board, Department of Science and Technology (SERB Grant No. EMR/2016/000362), to G. N.P. P. B. and N. K. S. acknowledge the computing facilities at the Indian Institute of Science Education and Research (IISER-Mohali). N. B. B. and N. K. S. thank CSIR India and DST-INSPIRE, respectively, for the research fellowship.

**Author contributions**

G.N.P. conceived and designed the experiment. N.B.B. performed all the experiments, in consultation with G.N.P. N.B.B. and G.N.P analyzed the data. N.K.S. carried out all the electronic structure calculations consultation with B.P. N.B.B., N.K.S, B.P., and G.N.P. discussed the results and contributed to the writing of the manuscript.

**Competing interests**

The authors declare no competing interests.

**Data availability**

The data that supports the plots within this paper and other findings of this study along with the methodology are available in the Supplementary Information.

# Supplementary Information

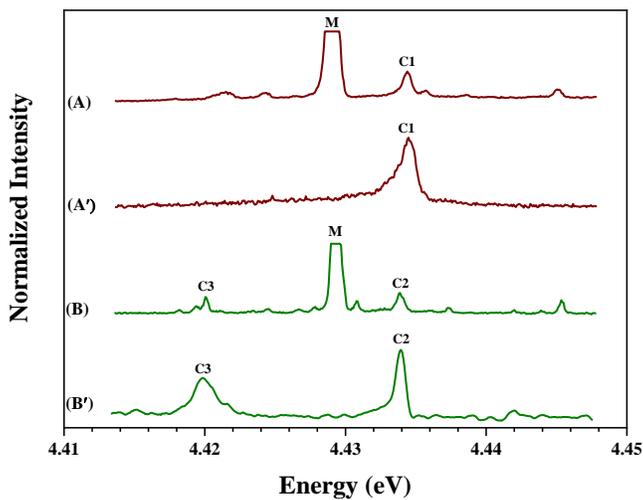

**Figure S1.** The electronic spectra of DFPHA complexes with DMA and TMA recorded using laser induced fluorescence (traces A, B) and resonant two-photon ionization method (traces A' and B'). In the spectrum recorded using laser induced fluorescence method, the band marked with M corresponds to the DFPHA monomer. The observation of fluorescence rules out the excited state electron transfer process from the amine to the DFPHA, which would lead to fluorescence quenching.

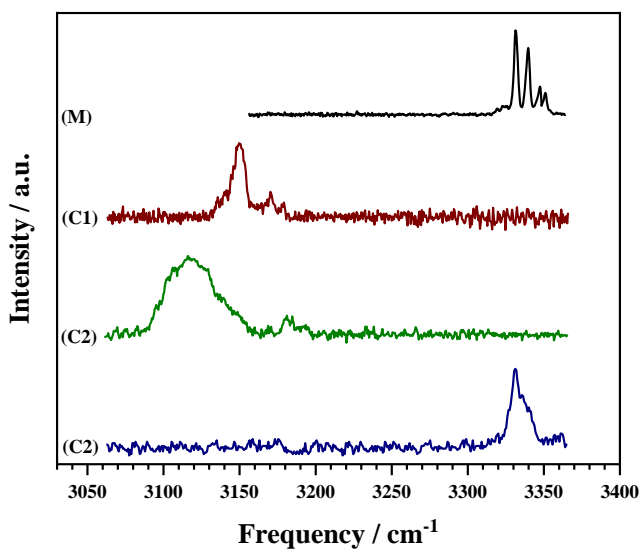

**Figure S2.** IR-UV double resonance spectra of the DFPHA monomer (M) and its complexes DMA (C1) and TMA (C2 and C3). The structures of the complexes shown in Figure 1 of the main text were inferred based on these spectra in combination with electronic structure calculations at MP2/cc-pVDZ level of theory.

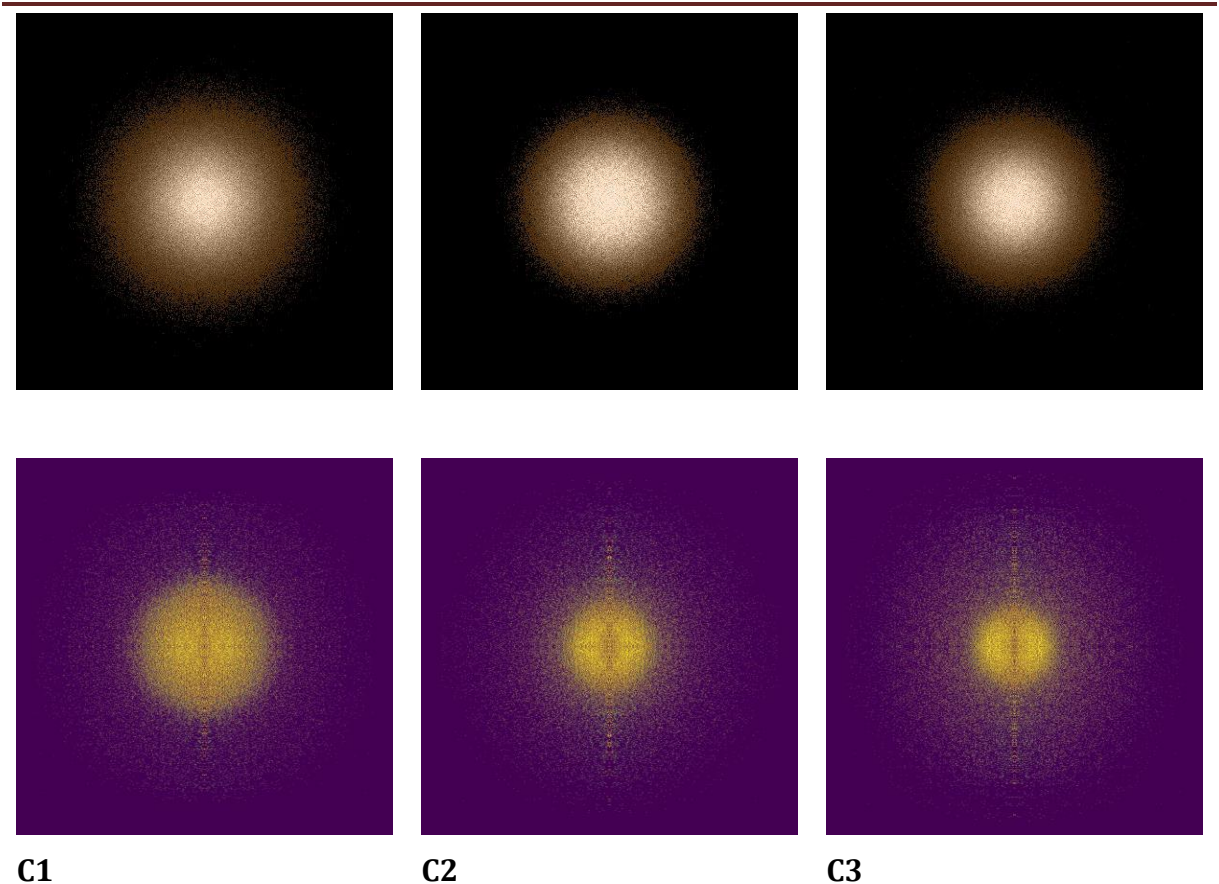

**Figure S3.** Raw and Abel-transformed velocity map images of the amine cations following resonant two-photon ionization of the corresponding binary complex with DFPHA recorded using centroiding mode

**Experimental Methodology**

The experiments were carried out on a skimmed molecular beam machine equipped with a velocity map imaging spectrometer.[1] Briefly, the binary complexes of 2,6-difluorophenylacetylene (DFPHA) with dimethylamine (DMA) / trimethylamine (TMA) were synthesized *in-situ* seeded in helium gas (4 bar). The methodology of the spectroscopic characterization of the binary complexes was reported earlier[2] and the amine cations were created following resonantly enhanced ionization of the binary complex with the excitation based on the DFPHA chromophore. The amine cations were collected using ion-optics which focused the ions onto a microchannel plate (MCP) detector coupled to a P47 phosphor screen (MCP-50DLP47VF; Tectra). The voltages on the lenses were adjusted such that the fragments with the same recoil velocity in the plane perpendicular to the flight direction are focused to the same point on the detector. The entire ion cloud was selectively detected by gating the front plate of the MCP detector by 75 ns high-voltage pulse (HTS 40-06-OT-75; Behlke) and the ensuing secondary electrons scintillate on the phosphor screen which was recorded by a CMOS camera (IDS-GigE). The images of amine cations were acquired both in the raw (20,000 laser shots) and centroiding (50,000 laser shots) modes with NuAcq software[3] and were Quadrant symmetrized[4] using ImageJ software.[5] Abel inversion was carried out by Basis Set Expansion method (BASEX) to extract the kinetic energy spectrum.[6]

**Computational Methodology**

The $C_{AC}$–H⋯N hydrogen-bonded complexes of DFPHA with DMA and TMA were optimized by varying the H⋯N hydrogen-bonded distance with a step size of 0.02 Å using Hartree-Fock (HF) level of theory with augmented correlation consistent-polarized valence double zeta (aug-cc-pVDZ) using Gaussian 09 suite of programs.[7] One dimensional potential energy curves (PECs) for the ground and the excited states of the neutral complexes and the ground state of the cationic complexes were calculated using Multi-Configuration Self-Consistent Field (MCSCF) level of theory with aug-cc-pVDZ basis set using a*b-initio* programs of MOLPRO package.[8] For MCSCF calculation, the set of frozen, closed orbitals and occupied orbitals are specified after obtaining it from HF calculation of the minimum ground state geometry. The list of molecular orbitals used for the DFPHA⋯DMA and DFPHA⋯TMA heterodimers and the corresponding monomers are shown in Figures S4-S7.

For the neutral DFPHA⋯DMA complex ($C_s$ symmetry) the ground state and the excited states were calculated with (34,22) set of occupied, (27,17) set of closed and (20,10) set of frozen orbitals as input for MCSCF calculations. This makes 8 electrons with 12 orbitals in the active space. The active space consisted of 3 occupied orbitals (HOMO, HOMO-2, HOMO-4) and 4 virtual orbitals (LUMO, LUMO+3, LUMO+4, LUMO+5) of A' symmetry, 1 occupied orbital (HOMO-1) and 4 virtual orbitals (LUMO+1, LUMO+2, LUMO+8, LUMO+11) of A'' symmetry. On the other hand, for the neutral DFPHA⋯TMA complex of $C_s$ symmetry, ground state and excited states were calculated with (37,23) set of occupied, (30,18) set of closed and (23,11) set of frozen orbitals as input for MCSCF calculations. This makes 8 electrons with 12 orbitals in the active space. The active space consisted of 3 occupied orbitals (HOMO,

**13**

HOMO-2, HOMO-4) and 4 virtual orbitals (LUMO, LUMO+1, LUMO+3, LUMO+5) of A' symmetry, 1 occupied orbital (HOMO-1) and 4 virtual orbitals (LUMO+2, LUMO+4, LUMO+7, LUMO+9) of A'' symmetry. The sets of occupied orbitals (From HOMO-4 to HOMO) show localization mostly over the benzene ring and the acetylenic moiety of DFPHA. There also exist lone pair like orbital on amine with HOMO-2 orbital. The choice of orbitals is such that it is localized over the acetylenic moiety of DFPHA and nitrogen atom of amine which will provide the necessary correlation energy correction in the calculation of excited state potential energy curves to reveal the relevant features.

To calculate exit channels of the excited state of DFPHA⋯DMA complex, the states of monomers DFPHA and DMA) were also calculated. For the monomer of neutral DFPHA ($C_{2v}$ symmetry), the ground state and the excited states were calculated with (20,6,13,4) set of occupied, (17,3,10,1) set of closed and (17,3,10,1) set of frozen orbitals as input for MCSCF calculations (8 electrons in 12 orbitals). For the monomer of neutral DMA and DMA$^+$ (both $C_s$ symmetry), the ground state and the excited states were calculated with (11,8) set of occupied, (6,3) set of closed and (6,3) set of frozen orbitals as input for MCSCF calculations (8 electrons in 10 orbitals). The excited states of monomers were used to characterize the exit channels. Similarly, to calculate exit channels of the excited states of DFPHA⋯TMA complex, the states of monomers (DFPHA and TMA) were also calculated. For the monomer of neutral DFPHA ($C_{2v}$ symmetry), the ground state and the excited states were calculated with (20,6,13,4) set of occupied, (17,3,10,1) set of closed and (17,3,10,1) set of frozen orbitals as input for MCSCF calculations (8 electrons in 12 orbitals). For the monomer of neutral TMA and TMA$^+$ ( both $C_s$ symmetry), the ground state and the excited states were calculated with (14,9) set of occupied, (9,4) set of closed and (9,4) set of frozen orbitals as input for MCSCF calculations (8 electrons in 10 orbitals). The excited states of monomers were used to characterize the exit channels.

For the cationic complex [DFPHA ⋯ DMA$^+$] of C$_S$ symmetry, the ground state was calculated with (34,22) set of occupied, (27,17) set of closed and (20,10) set of frozen orbitals as input for MCSCF calculations using ab-initio programs of MOLPRO. This makes 7 electrons with 12 orbitals in the active space. The active space consisted of 3 occupied orbitals (HOMO, HOMO-3, HOMO-4) and 4 virtual orbitals (LUMO, LUMO+1, LUMO+3, LUMO+4) of A' symmetry, 1 occupied orbital (HOMO-1) and 4 virtual orbitals (LUMO+2, LUMO+5, LUMO+6, LUMO+10) of A'' symmetry. On the other hand, for the cationic complex [DFPHA···TMA⁺] complex of $C_s$ symmetry, the ground state was calculated with (37,23) set of occupied, (30,18) set of closed and (23,11) set of frozen orbitals as input for MCSCF calculations. This makes 7 electrons with 12 orbitals in the active space. The active space consisted of 3 occupied orbitals (HOMO, HOMO-2, HOMO-4) and 4 virtual orbitals (LUMO, LUMO+3, LUMO+4, LUMO+5) of A' symmetry, 1 occupied orbital (HOMO-1) and 4 virtual orbitals (LUMO+1, LUMO+2, LUMO+9, LUMO+11) of A'' symmetry.

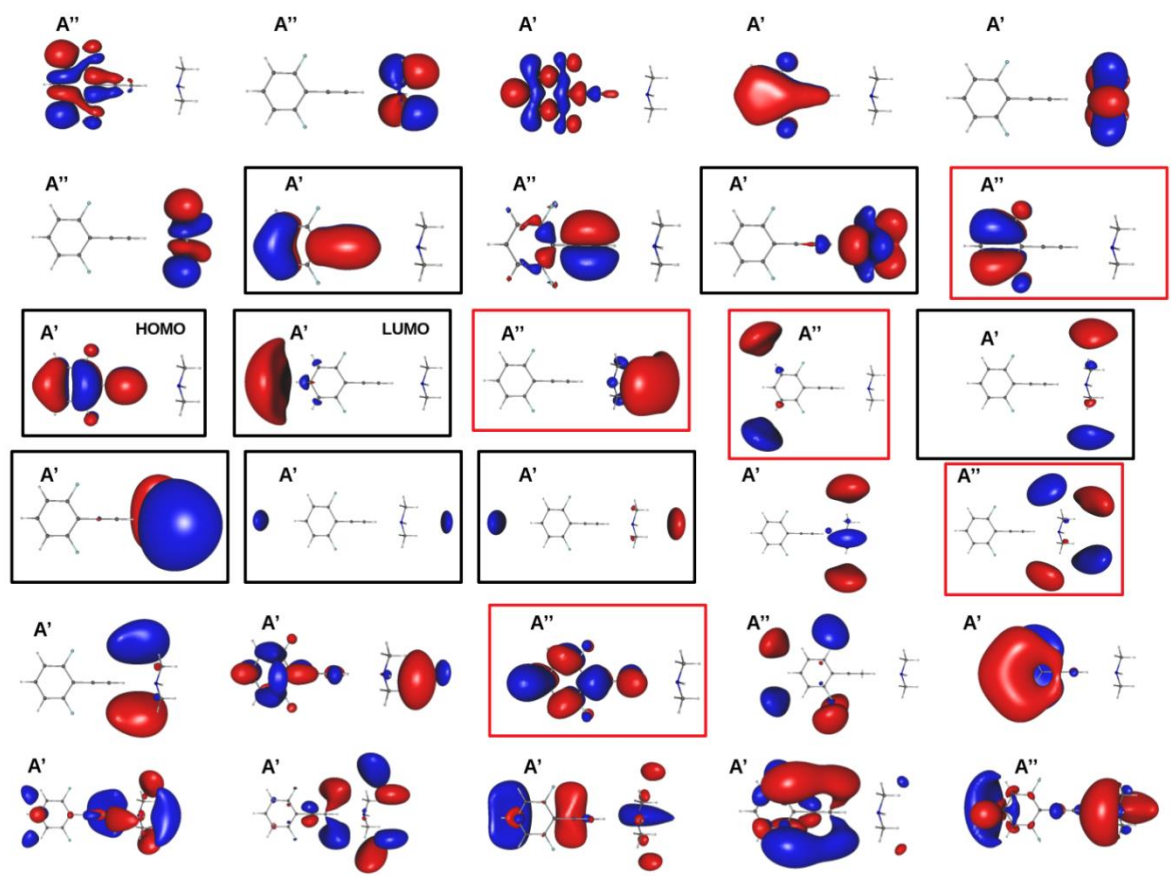

**Figure S4.** List of molecular orbitals plots obtained from HF/aug-cc-pVDZ calculation for the neutral DFPHA···DMA complex. The choice of active set for MCSCF calculation is made from visualizing these orbitals. The orbitals chosen to calculate A' and A'' states are enclosed inside the black and red rectangular boxes. The molecular orbitals surface is plotted at isovalue of 0.02 a.u.

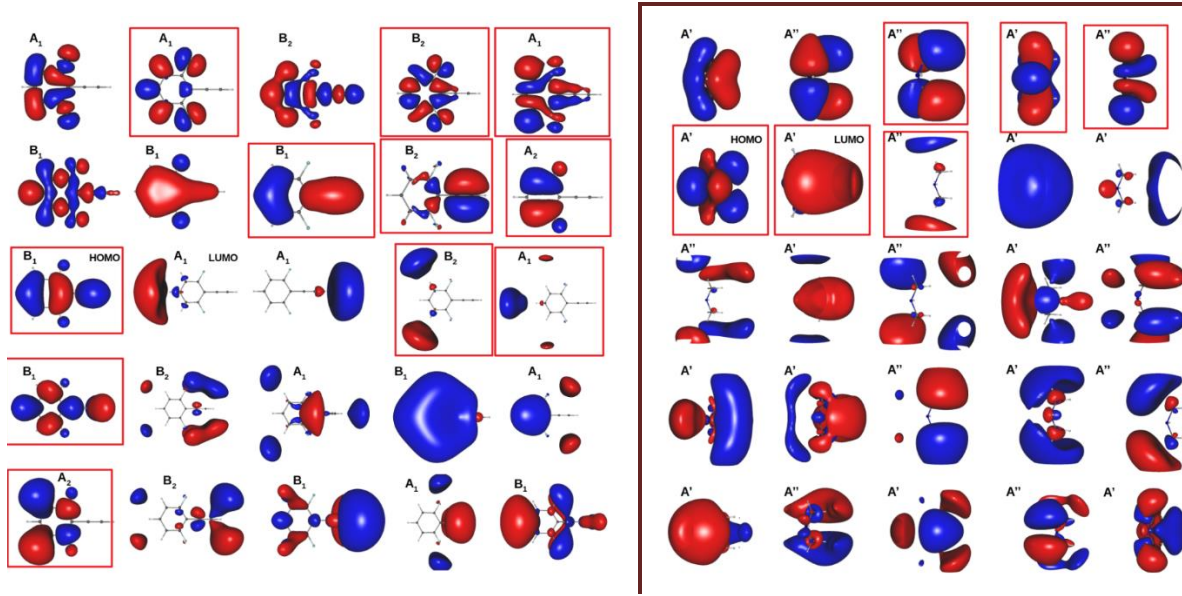

**Figure S5.** List of molecular orbitals plots obtained from HF/aug-cc-pVDZ calculation for the neutral monomers DFPHA ($C_{2v}$; left) and DMA ($C_s$; right) The choice of active set for MCSCF calculation is made from visualizing these orbitals. The orbitals chosen to calculate the excited states are enclosed inside red rectangular boxes. The molecular orbitals surface is plotted at isovalue of 0.07 a.u.

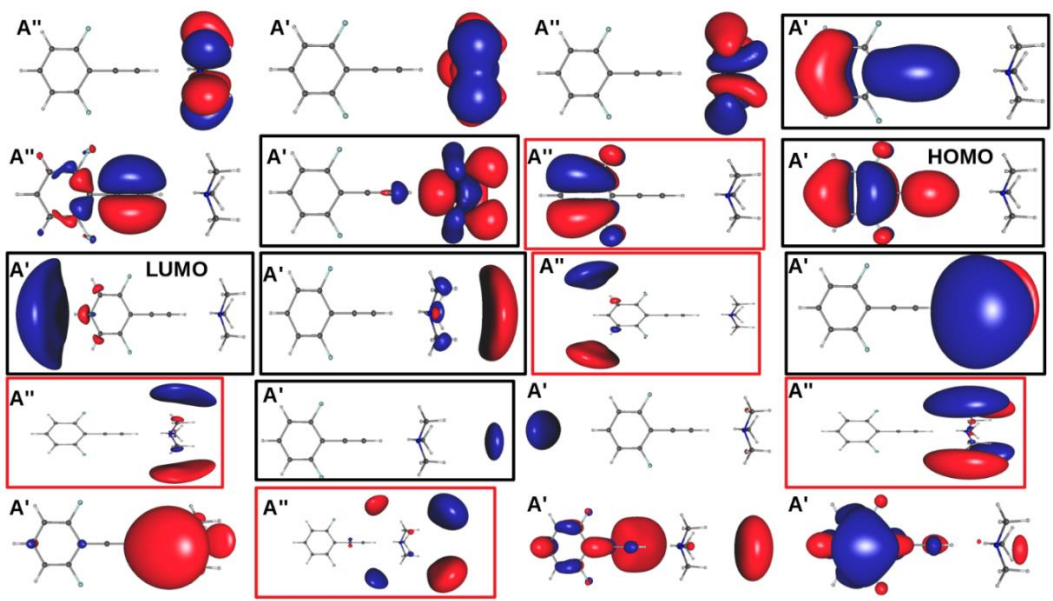

**Figure S6.** List of molecular orbitals plots obtained from HF/aug-cc-pVDZ calculation for the neutral DFPHA⋯TMA complex. The choice of active set for MCSCF calculation is made from visualizing these orbitals. The orbitals chosen to calculate A' and A'' states are enclosed inside the black and red rectangular boxes. The molecular orbitals surface is plotted at isovalue of 0.02 a.u.

**Figure S7.** List of molecular orbitals plots obtained from HF/aug-cc-pVDZ calculation for the neutral monomers DFPHA ($C_{2v}$; left) and TMA ($C_S$; Right). The choice of active set for MCSCF calculation is made from visualizing these orbitals. The orbitals chosen to calculate the excited states are enclosed inside red rectangular boxes. The molecular orbitals surface is plotted at isovalue of 0.07 a.u.

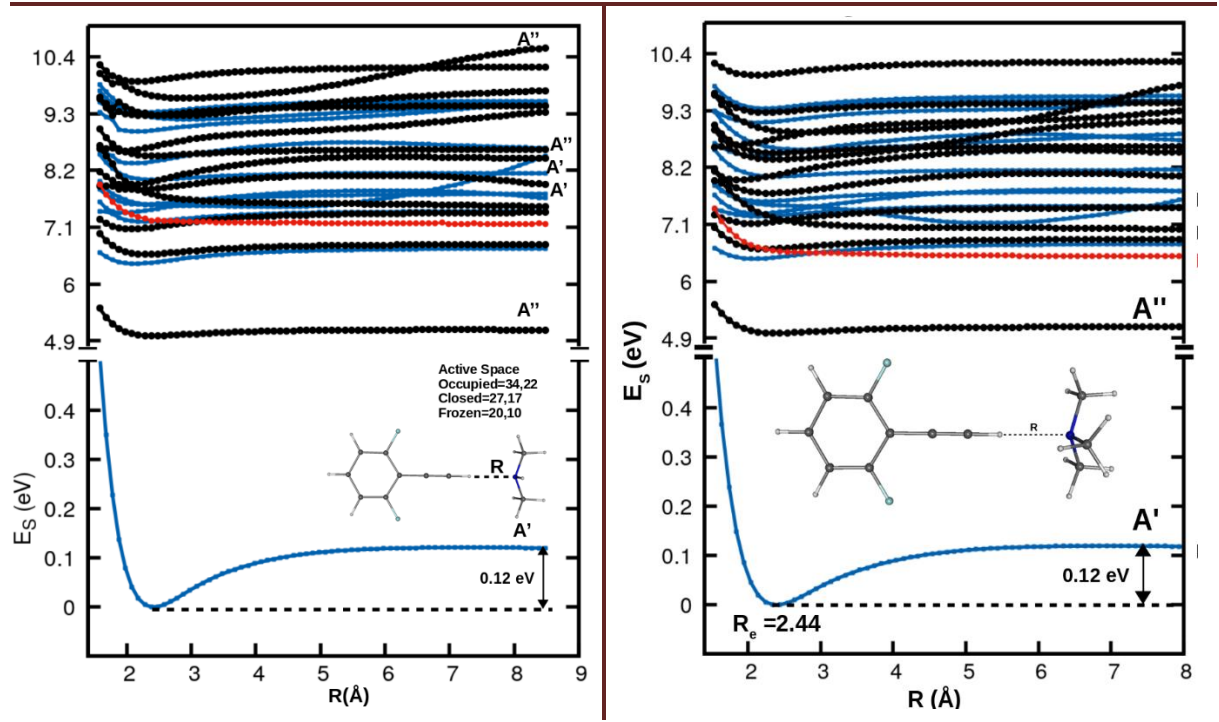

**Figure S8.** Relaxed PECs of the ground state, excited states of the neutral complex, and cationic ground state of DFPHA⋯DMA (left) DFPHA⋯TMA (right) complexes as a function hydrogen-bonded (R[$C_{AC}$–H⋯N]) distance. For the neutral complexes the A' and A'' states are shown in blue and black, respectively. The red curve represents the A' ground state of the cationic complex. The data for the PEC plots is given in Tables S1 and S2. The PECs are plotted relative to the minimum of the ground state.

The PECs of the neutral ground and excited states and the cationic ground states shown in Figure S8. The PEC plots shown in Figure 3 (of the main article) were obtained by shifting the excited states relative to the ground state such that the S1 state is at 4.43 eV.[2] On the other hand, the PEC of the cationic ground state was shifted such that the asymptote converges to the known ionization energy of neutral DMA (8.32 eV) and TMA (7.82 eV).[2,9]